\UseRawInputEncoding
\documentclass[superscriptaddress, twocolumn, amsmath, amssymb, aps,prl, letterdraft, notitlepage,longbibliography]{revtex4-1}
\usepackage{graphicx,graphics,epsfig,subfigure,times,bm,bbm,amssymb,amsmath,amsfonts,amsthm,mathrsfs,MnSymbol}
\usepackage[matrix,frame,arrow]{xypic}
\usepackage[normalem]{ulem}
\usepackage{slashed}
\usepackage{color}
\usepackage[usenames,dvipsnames,svgnames,table]{xcolor}
\usepackage[english]{babel}
\usepackage{verbatim}
\usepackage{tabularx}
\definecolor{darkblue}{rgb}{0.0,0.0,0.3}
\usepackage[colorlinks=true,
            linkcolor=red,
            urlcolor= darkblue,
            citecolor=blue]{hyperref}

\newcommand{\bea}{\begin{eqnarray}}
\newcommand{\eea}{\end{eqnarray}}

\begin{document}
\title{Boosting quantum battery performance by structure engineering}

\author{Junjie Liu}
\address{Department of Chemistry and Centre for Quantum Information and Quantum Control,
University of Toronto, 80 Saint George St., Toronto, Ontario, M5S 3H6, Canada}
\author{Dvira Segal}
\address{Department of Chemistry and Centre for Quantum Information and Quantum Control,
University of Toronto, 80 Saint George St., Toronto, Ontario, M5S 3H6, Canada}
\address{Department of Physics, 60 Saint George St., University of Toronto, Toronto, Ontario, M5S 1A7, Canada}

\begin{abstract}
Quantum coherences, correlations and collective effects can be harnessed to the advantage of
 quantum batteries.
Here, we introduce a feasible structure engineering scheme that is applicable to spin-based 
open quantum batteries. Our scheme, which builds solely upon a modulation of spin energy gaps, 
allows engineered quantum batteries to exploit spin-spin correlations for 
mitigating environment-induced aging. 
As a result of this advantage, an engineered quantum battery can preserve relatively more energy as 
compared with its non-engineered counterpart over the course of the storage phase. 
Particularly, the excess in stored energy is independent of system size.
This implies a scale-invariant passive protection strategy, which we demonstrate on an engineered 
quantum battery with staggered spin energy gaps. 
Our findings establish structure engineering as a useful route for advancing
quantum batteries, and bring new perspectives on efficient quantum battery designs.
\end{abstract}

\date{\today}

\maketitle

{\it Introduction.--}
Devising and realizing quantum batteries (QBs) \cite{Campaioli.18.NULL} 
is a rapidly growing research endeavour, requiring sustained and concerted efforts in 
quantum thermodynamics, quantum information, statistical mechanics, 
as well as atomic, molecular and optical physics to succeed. 
In this respect, numerous theoretical architectures are currently being pursued 
\cite{Alicki.13.PRE,Binder.15.NJP,Campaioli.17.PRL,Ferraro.18.PRL,Le.18.PRA,Andolina.18.PRB,Liu.19.JPCC,Santos.19.PRE,Andolina.19.PRL,Zhang.19.PRE,Pirmoradian.19.PRA,Andolina.19.PRBa,Farina.19.PRB,Rossini.20.PRL,Rossini.19.PRB,Barra.19.PRL,Hovhannisyan.20.PRR,Gherardini.20.PRR,Santos.20.PRE,Quach.20.PRA,BaiS.20.PRA,Rosa.20.JHEP,Kamin.20.NJP,Mitchison.20.NULL,Ghosh.20.PRA,Caravelli.20.PRR} (see Ref. \cite{Bhattacharjee.20.NULL} for a recent review). Among them, spin-based QBs \cite{Binder.15.NJP,Ferraro.18.PRL,Le.18.PRA,Andolina.18.PRB,Rossini.19.PRB,Andolina.19.PRL,Zhang.19.PRE,Pirmoradian.19.PRA,Andolina.19.PRBa,Rossini.19.PRB,Santos.20.PRE,Quach.20.PRA,BaiS.20.PRA,Ghosh.20.PRA,Caravelli.20.PRR,Kamin.20.NJP} represent arguably the most promising route towards applications, since spins can be realized in versatile contexts 
ranging from cavity/circuit quantum electrodynamics (QED) to solid state physics 
(see Refs. \cite{Blais.20.NP,Burkard.20.NRP} for reviews). The first experimental demonstration of spin-based QB has been carried out recently \cite{Quach.20.NULL}.

To date, a consensus has been reached that quantum mechanical resources 
such as entanglement and correlations are crucial for achieving quantum advantage of QBs. 
However, demonstrations of the quantum advantage of QBs are largely focused on the
charging and discharging stages \cite{Alicki.13.PRE,Campaioli.17.PRL,Le.18.PRA,Ferraro.18.PRL,Andolina.18.PRB,Andolina.19.PRL,Andolina.19.PRBa,Rossini.19.PRB,Zhang.19.PRE,Rossini.20.PRL,Rosa.20.JHEP,Seigi.20.PRR,Pintos.20.PRL}.
In comparison, obtaining a quantum advantage in the storage stage received 
less attentions due to the viewpoint that QBs in the storage stage can be treated as 
closed quantum systems that conserve energy 
(see, e.g., Ref. \cite{Ferraro.18.PRL}). 
However, it is now recognized that QBs in their storage stage warrant an open-system 
treatment \cite{Liu.19.JPCC,Santos.19.PRE,Rossini.19.PRB,Quach.20.PRA,Rosa.20.JHEP,Gherardini.20.PRR,BaiS.20.PRA,Santos.20.PRE,Mitchison.20.NULL} 
since their ability to store and conserve energy for later purposes is 
plagued by the unavoidable dissipation stemming from
interactions with surrounding environments. 
Hence, safeguarding open QBs during their storage phase is 
central to the mission of realizing QBs for quantum technological applications,
calling for efforts to exploit nontrivial quantum
effects for circumventing this practical challenge.

In this work, we introduce a simple and feasible route for harnessing spin-spin correlations 
for protecting spin-based QBs during the storage phase. 
Our proposal relies on a structure engineering (SE) of spin-based QBs by the
 modulation of spins energy gap, aiming for breaking the translational invariance of 
the bulk of spin-based QBs. 
Building on SE, spin-spin correlations are utilized to impact spin population dynamics, 
altering its decaying pattern from a fast exponential trend to a slower non-exponential decay. 
%
Spin-spin correlations can therefore be harnessed for mitigating aging of 
spin-based QBs \cite{Pirmoradian.19.PRA}, prolonging the storage time of charged QBs. 
The so-obtained protection strategy expands the family of passive protection protocols 
in the storage stage \cite{Liu.19.JPCC,Santos.19.PRE,Rossini.19.PRB,Quach.20.PRA,Rosa.20.JHEP}. 
We remark that passive protection strategies are favored from a thermodynamic perspective 
since their active counterparts (see Refs. \cite{Gherardini.20.PRR,BaiS.20.PRA,Santos.20.PRE,Mitchison.20.NULL}) cost extra energy for implementation, thereby reducing the overall efficiency of QBs. 

\begin{figure}[tbh!]
 \centering
\includegraphics[width=0.65\columnwidth] {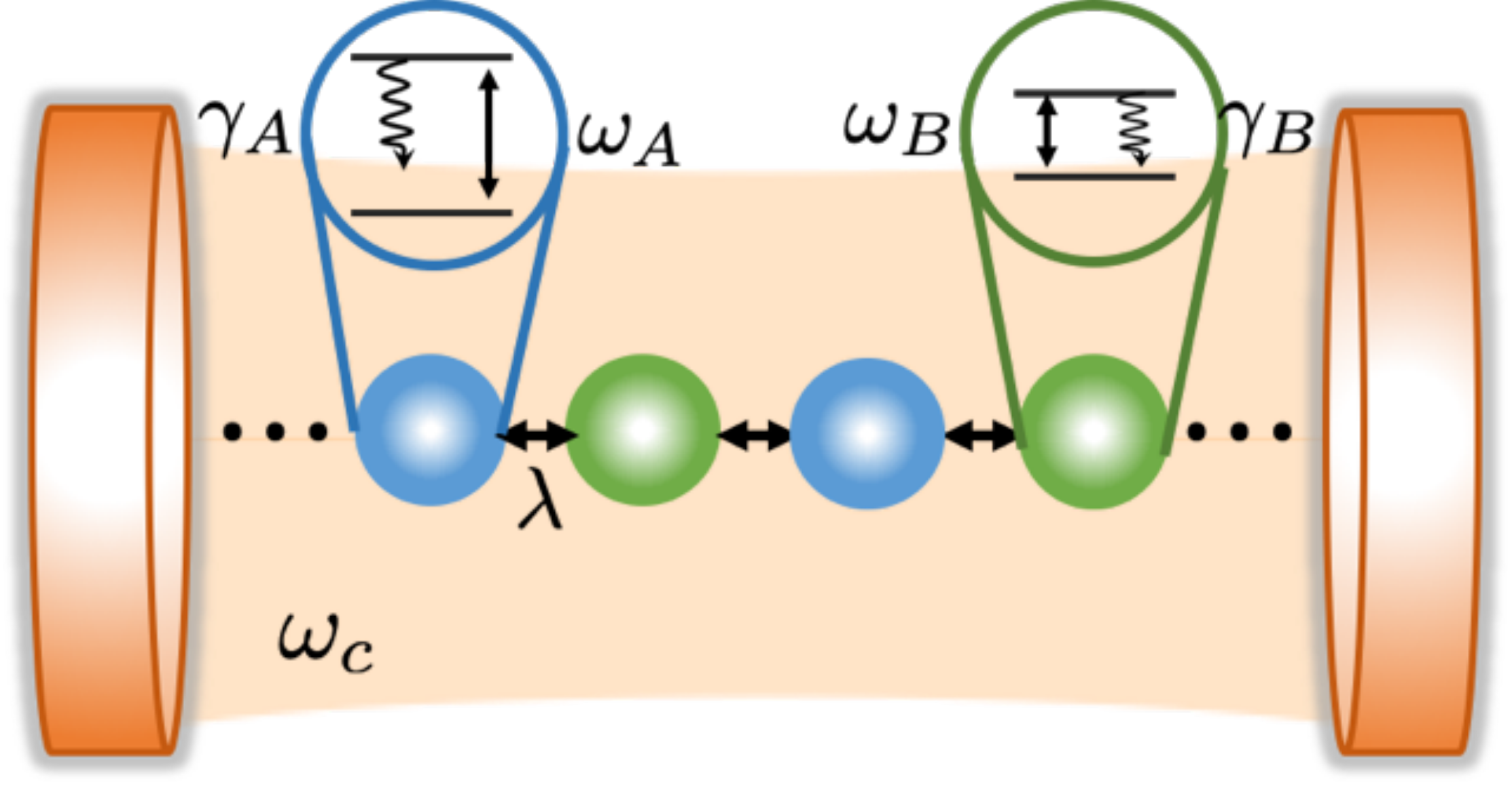}
\caption{Scheme of a structure-engineered quantum battery architecture consisting a cavity and a working 
substance--a dimeric spin lattice with staggered energy gaps $\omega_{A,B}$ and decay 
rates $\gamma_{A,B}$. Nearest-neighbor dipole-dipole couplings $\lambda$ is assumed uniform. 
In the storage phase, the dimeric spin lattice is decoupled from the cavity 
(frequency $\omega_c$),  
whereas in the charging and discharging phases the light-matter interaction is turned 
on and the cavity serves as a charger \cite{Ferraro.18.PRL,Andolina.18.PRB,Andolina.19.PRL,Quach.20.NULL} 
(with possible external driving fields) 
and a superradiant decay channel for spins \cite{Pirmoradian.19.PRA}, respectively.}
\label{fig:battery}
\end{figure}

To illustrate the SE strategy, we consider a prototype spin-cavity architecture 
for spin-based QBs \cite{Ferraro.18.PRL,Andolina.18.PRB,Andolina.19.PRL,Pirmoradian.19.PRA,Quach.20.NULL}. The working substance consists of a one-dimensional spin-$1/2$ lattice with uniform nearest-neighbor dipole-dipole coupling strengths. 
The potential of SE  is highlighted by contrasting a dimeric engineered QB with staggered 
spin energy gaps (see Fig. \ref{fig:battery}) to a non-engineered model with identical units.
%
%
Exploiting analytical solutions of quantum Lindblad master equation 
\cite{Lindblad.76.CMP} available for the single-excitation sector of the Hilbert space, 
we confirm the cooperation of spin-spin correlations in spin population dynamics 
in an engineered QB, leading to a slower spin decaying pattern. 
On the contrary, spin population dynamics in the non-engineered counterpart exhibits a fast 
exponential decay trend.
 From the energetics, quantified by the stored total energy, 
we further show that an engineered QB significantly outperforms its non-engineered counterpart, 
with more energy preserved during the storage phase. 
In particular, we find that the relative energy gain is independent of system size, 
indicating that we achieved a scale-invariant protection 
with potential applications to future scalable quantum battery setups.

{\it Microscopic model.--}We consider a spin-cavity 
architecture (see Fig. \ref{fig:battery} for a sketch) in light of existing 
theoretical proposals \cite{Ferraro.18.PRL,Andolina.18.PRB,Andolina.19.PRL,Pirmoradian.19.PRA}, 
and a recent experimental work \cite{Quach.20.NULL}.
The QB model includes a one-dimensional spin-$1/2$ lattice 
as the working substance with spin gaps $\{\omega_j\}$, 
Pauli spin operators 
$\{\sigma_j^{z,\pm}\}$ ($\sigma_j^-=(\sigma_j^+)^{\dagger}$), $j\in[1,\cdots,N]$ (setting $N$ an even number), and nearest-neighbor dipole-dipole interactions measured by the strength $\lambda$. 
The spins are coupled to an optical cavity 
supporting a single dispersionless mode with frequency $\omega_c$ 
and annihilation operator $a$. The light-matter coupling, measured by strength $\eta$, 
is treated within the rotating-wave-approximation. 
In the rotating frame, we obtain the total Hamiltonian (setting $\hbar=1$),
\begin{equation}
H_{\mathrm{tot}}~=~\sum_{j=1}^N\frac{\Delta_j}{2}\sigma_j^z+\sum_{j=1}^{N-1}\lambda(\sigma_j^+\sigma_{j+1}^-+\mathrm{H.c.})+\eta\sum_{j=1}^N(\sigma_j^-a^{\dagger}+\mathrm{H.c.}).
\end{equation}
Here, $\Delta_j=\omega_j-\omega_c$, `H.c' is short for `Hermitian conjugate'. 
Our results in the storage phase do not depend on the frequency $\omega_c$ itself. 

In addition to the coherent Hamiltonian $H_{\mathrm{tot}}$, incoherent processes affect the QB.
These include the decay of the intensity of the field within the cavity, 
with a rate constant  $\kappa$,
and the spontaneous decay of spins, with rate constants $\gamma_j$. 
In the weak dissipation regime of $\gamma_j\ll\Delta_j$, $\kappa\ll\omega_c$, 
the total density matrix $\rho_{\mathrm{tot}}$ of the cavity and spins 
is governed by the following Lindblad master equation 
\cite{Lindblad.76.CMP} (explicit time dependence is suppressed)
\begin{equation}\label{eq:qme_tot}
\frac{\partial}{\partial t}\rho_{\mathrm{tot}}~=~-i[H_{\mathrm{tot}},\rho_{\mathrm{tot}}]+\kappa\mathcal{L}[a]\rho_{\mathrm{tot}}+\sum_{j=1}^N\gamma_j\mathcal{L}[\sigma_j^-]\rho_{\mathrm{tot}}.
\end{equation}
Here, $\mathcal{L}[O]\rho_{\mathrm{tot}}=(2O\rho_{\mathrm{tot}}O^{\dagger}-O^{\dagger}O\rho_{\mathrm{tot}}-\rho_{\mathrm{tot}}O^{\dagger}O)/2$ denotes the Lindblad superoperator. 
We neglect dephasing and heating (incoherent pumping) of spins, as their rates can be 
made several orders of magnitude smaller than individual decay rates \cite{Schwager.13.PRA}. 

We limit our analysis to the storage phase. 
Following Ref. \cite{Pirmoradian.19.PRA} we assume that the QB is fully charged 
(through the cavity \cite{Ferraro.18.PRL,Andolina.18.PRB,Andolina.19.PRL}) at 
time $t=0$ when the storage stage begins. During the storage stage we turn off the light-matter 
interaction by tuning the cavity frequency away from $\omega_{j}$. 
Hence, in the storage stage we just consider the reduced quantum master equation for 
spins (see the supplemental material \cite{SM} for more details),
\bea\label{eq:rho_ss}
\frac{\partial}{\partial t}\rho_{S} &=& -i[H_S,\rho_S]+\sum_{j=1}^N\gamma_j\mathcal{L}[\sigma_j^-]\rho_S~\equiv~\mathbb{L}\rho_S.
\eea
Here, $H_S=\sum_{j=1}^N\frac{\Delta_j}{2}\sigma_j^z+\sum_{j=1}^{N-1}\lambda(\sigma_j^+\sigma_{j+1}^-+\sigma_{j+1}^+\sigma_{j}^-)$ 
is the spin lattice Hamiltonian in the rotating frame, $\mathbb{L}$ is the total Liouvillian. 

{\it Structure engineering scheme.--} 
We begin with dynamical equations for spin population $\langle\sigma_z^j\rangle$ 
obtained using Eq. (\ref{eq:rho_ss}),
\bea\label{eq:sigz_eom}
\frac{d}{dt}\langle \sigma_j^z\rangle &=& -2i\lambda\Big[\langle \sigma_j^+\sigma_{j+1}^-\rangle+\langle\sigma_j^+\sigma_{j-1}^-\rangle-\langle\sigma_{j+1}^+\sigma_j^-\rangle\nonumber\\
&&-\langle\sigma_{j-1}^+\sigma_j^-\rangle\Big]-\gamma_j\langle 1+\sigma_j^z\rangle.
\eea
For long spin lattices with $N\gg1$, we can neglect boundary effects 
and focus on the bulk of the QB. 
We observe that the first four terms, 
representing spin-spin correlations, cancel out exactly in the bulk of a uniform spin
lattice with a translational 
invariance since, for instance, $\langle \sigma_{j-1}^+\sigma_{j}^-\rangle=\langle \sigma_j^+\sigma_{j+1}^-\rangle$. 
As a result, even though spin-spin correlations are nonzero, 
the spin population of a uniform spin lattice decays exponentially and locally, 
in the sense that the decaying dynamics is 
independent of other spins, leading to an aging of charged QBs \cite{Pirmoradian.19.PRA}.

Can one harness spin-spin correlations to mitigate aging? 
We answer this question affirmatively by introducing a strategy based on SE, 
building solely on the modulation of spin energy gaps such that, 
for instance, we should at least have $\Delta_j\neq\Delta_{j\pm1}$ in the present model. 
%
Through SE, realized here with the addition of  
non-homogeneity of spin energy gaps to the otherwise uniform lattice, we 
break the translation invariance of the bulk such that 
$\langle \sigma_{j-1}^+\sigma_{j}^-\rangle\neq\langle \sigma_j^+\sigma_{j+1}^-\rangle$, resulting in nonzero contributions from spin-spin correlations to the spin population dynamics.
The result of the inclusion of spin-spin correlations is that we can alter the decaying pattern 
of spin populations from a fast exponential trend to a non-exponential one. More intriguingly, the SE simultaneously modifies the spontaneous emission rate $\gamma_j$ as it is proportional to the spin energy gap (see, e.g., Ref. \cite{Lodahl.15.RMP}). Hence the engineered decay dynamics can become relatively slower, 
thereby mitigating aging effect in a passive manner. 

Although here we lay out the SE scheme by focusing on a spin lattice with nearest-neighbor couplings, 
we emphasize that our SE scheme is not limited to this specific case. 
In fact, SE strategies can be tailored for other spin lattice models by inspecting the detailed 
form of the dynamical equations governing spin population dynamics,
 which is model dependent; more discussions will be provided later.

{\it Dimeric lattice and single-excitation sector.--}To facilitate the analysis with numerical insights, 
we adopt a dimeric SE: A dimeric spin lattice
with staggered energy gaps $\Delta_{A,B}$ and decay rates $\gamma_{A,B}$ ($\gamma_A/\Delta_A=\gamma_B/\Delta_B$) but uniform dipole-dipole coupling strength 
$\lambda$ (see Fig. \ref{fig:battery}) \footnote{For systems with nearest-neighbor couplings, we note that a dimeric lattice is sufficient to break the translational invariance. However, one can consider more complicated structural engineering, 
for instance, a trimeric lattice.}. Without loss of generality, 
we set $\Delta_j=\Delta_{A(B)}$ if $j$ is odd (even). 
We note that a dimeric SE can either speed or slow the decay dynamics of the spin populations, 
compared to that of the non-engineered lattice ($\Delta_A=\Delta_B$), 
depending on the ratio $\Delta_{B}/\Delta_A$. 
For our purposes, we refer to the system with $\Delta_A=\Delta_B$ ($\Delta_A>\Delta_B$) as the non-engineered (engineered) QB.

We first focus on the single-excitation sector of the spin Hilbert space 
containing one spin excitation in total. 
Notably, the slowest decay dynamics of the system belongs to this sector 
\cite{Torres.14.PRA,Cabot.19.PRL}. 
Since the jump term $\sum_j\gamma_j\sigma_j^-\rho_S\sigma_j^+$ 
of the Lindblad master equation Eq. (\ref{eq:rho_ss}) does not contribute in this sector, 
the dissipative dynamics is fully governed by the part $\partial\rho_S/\partial t=\mathcal{K}\rho_S\equiv-i[K\rho_S-\rho_SK^{\dagger}]$ 
with an effective (non-Hermitian) Hamiltonian 
$K\equiv H_S-i\sum_j\frac{\gamma_j}{2}\sigma_j^+\sigma_j^-$.
Since $K$ is quadratic, one easily finds its 
two-band complex eigenvalues under an open boundary condition \cite{Cabot.19.PRL}
$\Omega_k^{\pm}~=~[(\Omega_A+\Omega_B)/2]\pm\frac{1}{2}\sqrt{(\Omega_A-\Omega_B)^2+16\lambda^2\cos^2(k/2)}$.
%
Here, $\Omega_{A(B)}=\Delta_{A(B)}-i\gamma_{A(B)}$ and $k=2\pi l/(N+1)$
 with $l=1,2,\cdots,N/2$. 
The eigenvalues of $K^{\dagger}$ are the complex conjugates $\Omega_k^{\pm\ast}$. 
In general, the eigenvalues of the Liouvillian $\mathbb{L}$ can be constructed by using 
$\Omega_k^{\pm}$ and $\Omega_k^{\pm\ast}$ as detailed in Ref. \cite{Torres.14.PRA}, 
while the smallest decay rates are $-\mathrm{Im}[\Omega_k^{\pm}]$ with `Im' taking imaginary part. 
For a uniform lattice with $\Omega_A=\Omega_B$, we have $-\mathrm{Im}[\Omega_k^{\pm}]=\gamma_A$. 
Notably, the inverse of $\gamma_A$ sets the storage timescale for the non-engineered QB 
\cite{Pirmoradian.19.PRA}. On the contrary, we find that the smallest decay rates of the dimeric spin
 lattice is smaller than that of the uniform lattice (see the supplemental material \cite{SM}), indicating a longer storage time.

To quantify the dynamical behavior, we study spin populations 
$\langle\sigma_j^{z}(t)\rangle$ and spin-spin correlations 
$\langle\sigma_j^+(t)\sigma_{j'}^-(t)\rangle$, which have the following analytical expressions in the single-excitation sector \cite{Cabot.19.PRL}, 
\bea\label{eq:sigma}
&&\langle\sigma_j^z(t)\rangle~=~2\sum_{n,m}G_{n,m}(j)e^{-i(\Omega_n-\Omega_{m}^{\ast})t}-1,\nonumber\\
&&\langle\sigma_j^+(t)\sigma_{j'}^-(t)\rangle~=~\sum_{n,m}W_{n,m}(j,j')e^{-i(\Omega_n-\Omega_{m}^{\ast})t}.
\eea
Here, $\Omega_n$ with $n$ running from $1$ to $N$ are elements of a $1\times N$ vector 
$\boldsymbol{\Omega}$ whose first (second) half belongs to $\Omega_k^{-}$ ($\Omega_k^{+}$) with $k$ running from $2\pi/(N+1)$ to $\pi N/(N+1)$ (see above).
The coefficients $G_{n,m}(j)$ and $W_{n,m}(j,j')$ are determined by the eigenvectors of 
$K$ and $K^{\dagger}$, as well as by the initial condition $\rho_S(0)$;
we relegate their detailed expressions to the supplemental material \cite{SM}.
 
\begin{figure}[tbh!]
 \centering
\includegraphics[width=1\columnwidth] {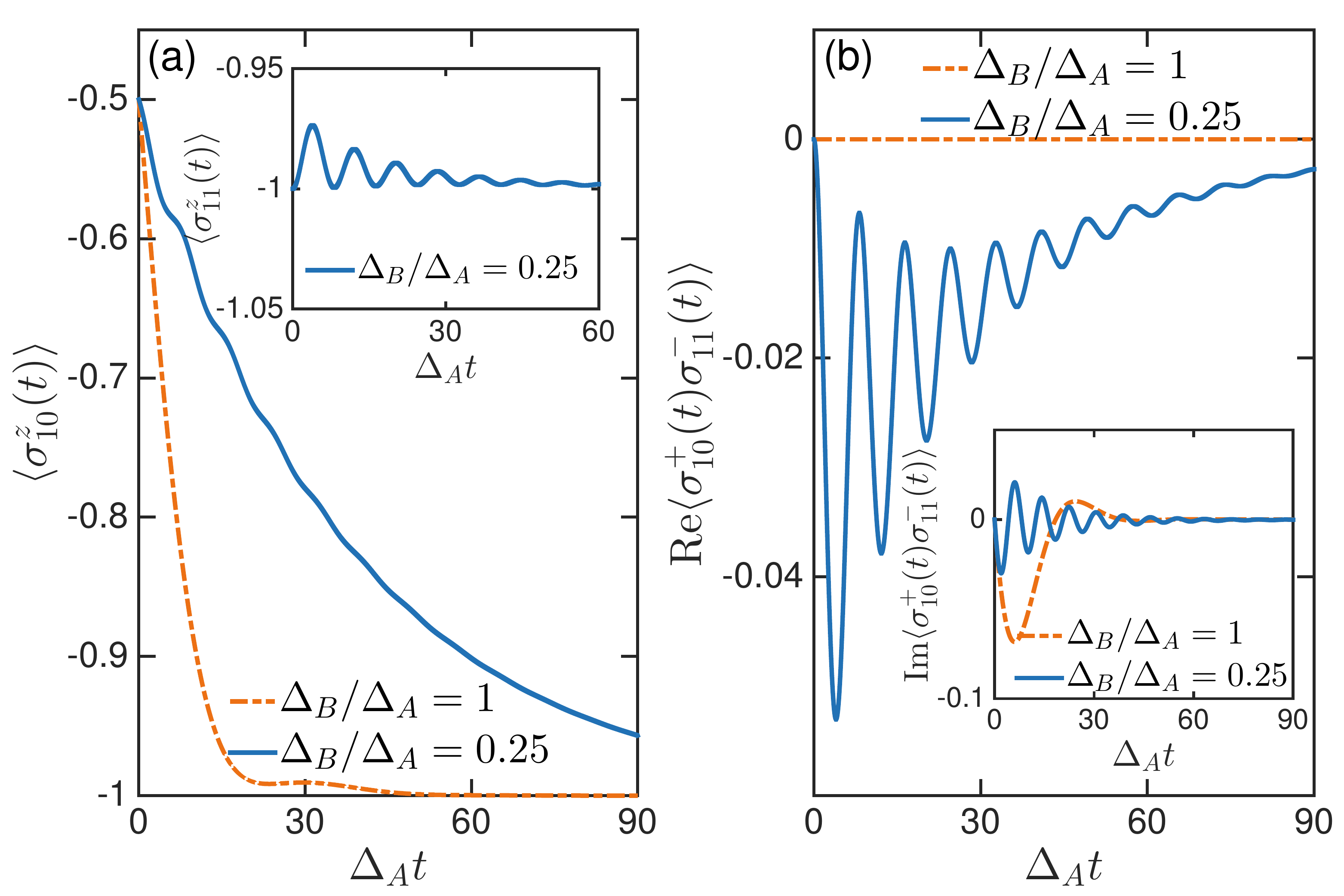}
\caption{(a) Trajectory of $\langle\sigma_{10}^z(t)\rangle$ in a single-excitation 
sector using Eq. (\ref{eq:sigma}) for $\Delta_B/\Delta_A=0.25$ (blue solid line) and $\Delta_B/\Delta_A=1$ (orange dash-dotted line).
Inset: Trajectory of $\langle\sigma_{11}^z(t)\rangle$ for $\Delta_B/\Delta_A=0.25$. 
(b) Trajectory of the real (Re) part of the spin-spin correlation 
$\langle\sigma_{10}^+(t)\sigma_{11}^-(t)\rangle$ using Eq. (\ref{eq:sigma}) for $\Delta_B/\Delta_A=0.25$ (blue solid line) and $\Delta_B/\Delta_A=1$ (orange dash-dotted line). 
Inset: Trajectory of the imaginary (Im) part of the spin-spin correlation $\langle\sigma_{10}^+(t)\sigma_{11}^-(t)\rangle$. Other parameters are $N=50$, $\gamma_{A,B}/\Delta_{A,B}=0.05$, $\lambda=0.05\Delta_A$.}
\label{fig:syn}
\end{figure}

In Fig. \ref{fig:syn}, we present results for population $\langle\sigma_j^{z}(t)\rangle$ 
and spin-spin correlation $\langle\sigma_j^+(t)\sigma_{j'}^-(t)\rangle$ at the bulk of
the spin chain (sites $j=10$ and $j=11$).
We use Eq. (\ref{eq:sigma}) and assume the (arbitrary) initial state 
$\rho_S(0)=|\Phi_0\rangle\langle\Phi_0|$; 
$|\Phi_0\rangle=|g\rangle/\sqrt{2}+(|e_{j=10}\rangle+|e_{j=12}\rangle)/2$  with 
$|e_j\rangle=\sigma_j^+|g\rangle$ 
and $|g\rangle$ the global ground state of the spins, 
namely, $\sigma_j^z|g\rangle=-|g\rangle$. 
We confirmed that basic features depicted in Fig. \ref{fig:syn} are independent of the 
initial condition adopted in the single-excitation sector.
%
%
From Fig. \ref{fig:syn} (a), it is evident that spin population in a uniform (non-engineered) lattice
 (dash-dotted line) decays initially in an exponential manner. 
In comparison, the spin population in a dimeric lattice (solid line) depicts a slower decay,
dressed by an oscillatory behavior at short times. 
Moreover, by the time $\Delta_A t=25$, the excited state population in the uniform case is close to zero,
while in the dimerized case by that time only 40$\%$ had decayed to the ground state. 

Based on the SE scheme and Eq. (\ref{eq:sigz_eom}), 
we naturally expect that transient oscillations of the spin population 
(see also the inset for a nearest-neighbor site) 
in the dimeric lattice arise from the spin-spin correlations. 
To verify whether this is the case, we turn to the spin-spin correlation result 
showed in Fig. \ref{fig:syn} (b). 
For clarity and simplicity, we just depict $\langle \sigma_{10}^+(t)\sigma_{11}^-(t)\rangle$ 
in accordance with population results of Fig. \ref{fig:syn} (a).
From Fig, \ref{fig:syn} (b), we immediately note that spin-spin correlations 
in the dimeric lattice oscillate with a period that is consistent with that inferred from spin 
population dynamics at short times, confirming that spin-spin correlations 
indeed affect spin population dynamics in the dimeric lattice. 
In comparison, although we have nonzero spin-spin correlations in a uniform lattice, 
its impact on spin population dynamics is negligible, in accordance with Eq. (\ref{eq:sigz_eom}).  
Hence, from the dynamics in the single-excitation sector, we confirm that SE (i) 
allows for the participation of spin-spin correlation in spin population dynamics,
 and (i) leads to slower decaying trend of spin populations, thereby mitigating the aging effect.

{\it Energetics in the storage phase.--}
To characterize the performance of a fully-charged engineered QB, we turn to the energetics 
in the storage phase. This  requires information of higher-order excitation sectors 
as we have $N$ excitations at $t=0$ 
($\langle \sigma_j^z(t=0)\rangle=1$ but spin coherences and correlations are set to zero), 
namely, the full Lindblad master equation Eq. (\ref{eq:rho_ss}) should be utilized. 
We consider the dynamics of the average total energy normalized by frequencies,
\begin{equation} 
\mathcal{E}(t)\equiv \frac{\langle H_S(t)\rangle}{\sum_{j=1}^N\Delta_j/2}.
\end{equation}
This measure allows for a proper comparison between engineered and non-engineered QBs. 
It has been demonstrated for spin-based QBs that the average total energy approaches 
the ergotropy--the maximal extractable work under a cyclic 
unitary transformation \cite{Allahverdyan.04.EPL}--in the large $N$ limit \cite{Rossini.19.PRB,Andolina.19.PRL,Quach.20.PRA}. We thus perform simulations in this limit.
We denote by $\mathcal{E}_{d(u)}(t)$ 
the average total energy in the dimeric (uniform) lattice. 
The dynamics of $\mathcal{E}(t)$ is obtained by solving Eq. (\ref{eq:sigz_eom}) 
together with those for higher-order correlation terms based on Eq. (\ref{eq:rho_ss}); 
coupled equations of motion obtained under a second-order cumulant approximation 
\cite{Holland.14.PRL,Meiser.10.PRAa} applicable for large $N$ are listed in the 
supplemental material \cite{SM}.

In Fig. \ref{fig:EE},  we show the dynamics of the relative energy excess 
$[\mathcal{E}_d(t)-\mathcal{E}_u(t)]/\mathcal{E}_u(t)$ during the storage phase; 
a similar comparison for the averaged population is depicted in the supplemental material \cite{SM}.
%
\begin{figure}[tbh!]
 \centering
\includegraphics[width=0.95\columnwidth] {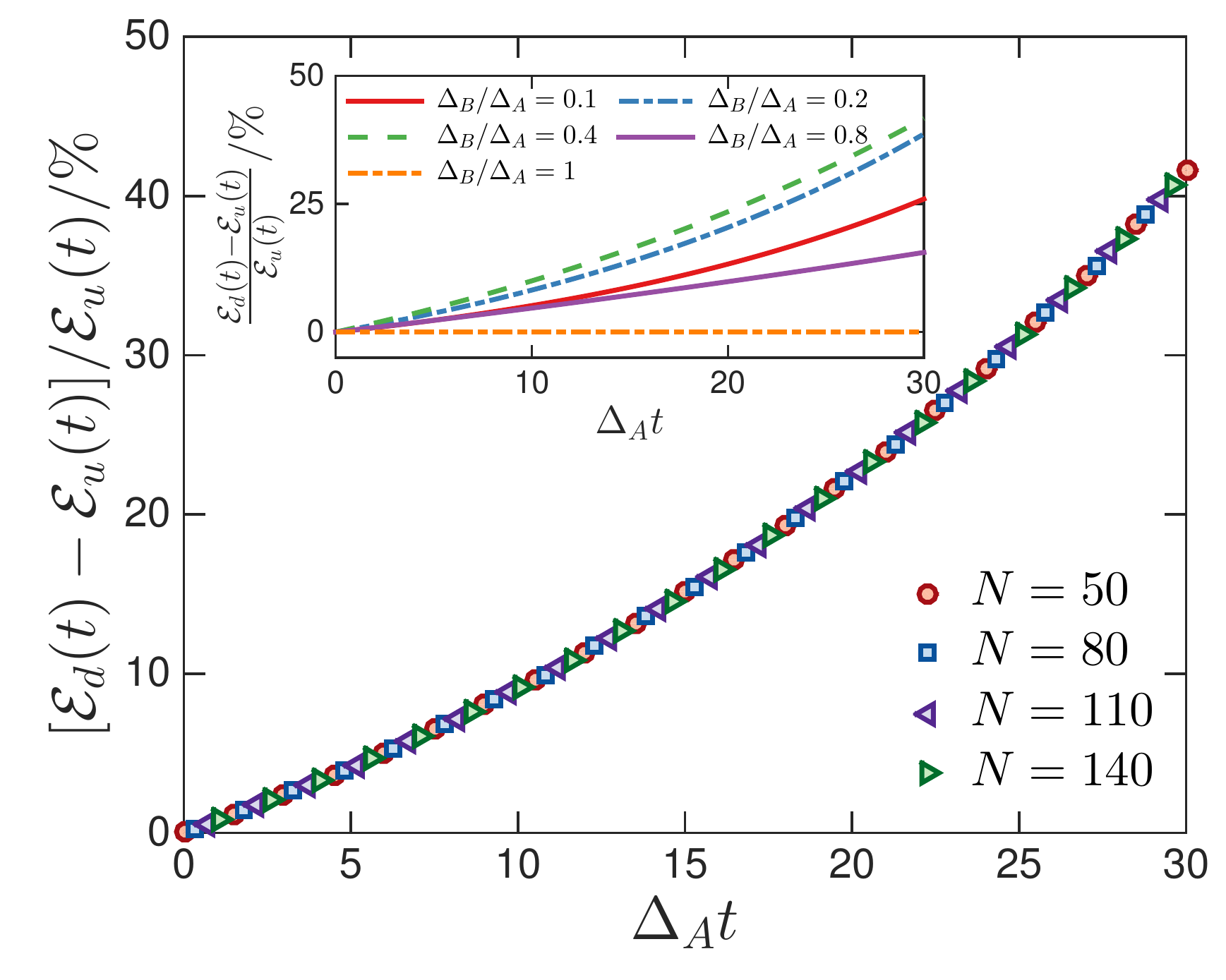}
\caption{Dynamics of relative stored energy excess $[\mathcal{E}_d(t)-\mathcal{E}_u(t)]/\mathcal{E}_u(t)$ in the storage phase as a function of spin numbers $N$ with a fixed detuning ratio $\Delta_B/\Delta_A=0.25$. 
Inset: Relative excess energy when varying detuning ratio $\Delta_B/\Delta_A$ at a fixed number $N=80$. 
Other parameters are $\gamma_{A(B)}/\Delta_{A(B)}=0.05$ and $\lambda=0.05\Delta_A$.}
\label{fig:EE}
\end{figure}
We observe several features that are worth mentioning: 
(i) The engineered QB preserves more energy than the non-engineered counterpart 
as the storage time goes. 
Particularly, the relative excess $[\mathcal{E}_d(t)-\mathcal{E}_u(t)]/\mathcal{E}_u(t)$ is
monotonic as a function of storage time. Approximately, we reveal the scaling 
$\mathcal{E}_d(t)/\mathcal{E}_u(t)\propto t^{\alpha}$
with $\alpha=1.5$ at long times \cite{SM}. 
Notwithstanding, we point out that the absolute excess 
$\mathcal{E}_d(t)-\mathcal{E}_u(t)$ shows a turnover behavior  
when increasing the storage time \cite{SM}, implying an optimal protection time for the engineered QB.  
Notably, one can also infer this property from Fig. \ref{fig:syn} (a) 
for the population dynamics in the one-excitation sector, 
identified by the maximum population contrast appears around $\Delta _A t \sim 30$. 
(ii) The {\it relative} excess is independent of the number of spins $N$, 
indicating that the resulting passive protection scheme is scale-invariant 
and can be applied to large-scale QBs.
 (iii) The inset shows that the relative excess is a nonmonotonic function in 
the frequency ratio $\Delta_B/\Delta_A$, with the maximum achieved at an intermediate value. 
This property indicates that the so-obtained advantage of engineered QBs is not merely a consequence 
of the modification of decay rates from
 $\gamma_A$ to $\gamma_B$ for spins with even indices, 
otherwise, we should have observed a monotonic dependence of 
the relative excess on the frequency ratio $\Delta_B/\Delta_A$ as $\gamma_B$ monotonically 
approaches $\gamma_A$ when we increase the ratio. 
We note that such a dimeric spin lattice allows for an intriguing 
collective spin motion, namely, quantum {\it transient} synchronization \cite{Cabot.19.PRL} 
as featured by decaying oscillations of spin coherences 
$\langle\sigma_j^x(t)\rangle$, with the same frequency, in spite of distinct intrinsic spin
frequencies (see the supplemental material \cite{SM} for trajectories 
with varying $\Delta_B/\Delta_A$). Interestingly, this synchronization only occurs 
within a specific range of frequency ratio \cite{Cabot.19.PRL,SM}. 
We therefore argue that the observed nonmonotonic behavior in
energy stored indicates that the nontrivial collective spin motion enabled by spin-spin correlations
contributes to the advantage of engineered QBs.

{\it Discussion.--}
We envision realization of the engineered QB using circuit QED architectures 
involving superconducting qubits 
\cite{Houck.12.NP,Salathe.15.PRX,Muller.19.RPP,Carusotto.20.NP,Blais.20.NP,Burkard.20.NRP},
leveraging an exquisite tunability over system parameters such as the qubit frequency.
Furthermore, one can easily scale up the system size in circuit QED  
\cite{Houck.19.N}, allowing for scalable QB designs.
The implementation of the engineered QB is not tied to circuit QED setups 
in light of parallel advances in a variety of other experimental platforms, 
including cold atoms \cite{Reitz.13.PRL,Goban.14.NC,Douglas.16.PRX,Bernien.17.N}, 
quantum dots \cite{Yalla.14.PRL,Arcari.14.PRL}, and trapped ions \cite{Barreiro.11.N} 
in optical traps and photonic structures, 
where spins connected in a one-dimensional arrangement are concerned.

Two possible extensions of the present strategy to more complicated QB designs are anticipated: 
(i) Many-body QBs with long-range spin-spin couplings 
beyond nearest-neighbor order \cite{Le.18.PRA,Rossini.19.PRB}.
In this scenario, equations of motion for spin populations involve long-range correlation terms and we expect engineered structures beyond a dimeric configuration, 
depending on the details of the underlying spin-spin interaction pattern. 
(ii) QBs in higher dimensions. Taking a possible two-dimensional QB as an example, 
we anticipate that the same approach of SE would directly apply if we limit spin-spin 
interactions to lowest nearest-neighbor terms. 
Although experimental techniques for connecting spins in two-dimensional arrangements 
are mature \cite{Bloch.08.RMP,Bloch.12.NP}, it is challenging to implement the dimeric 
lattice as it creates a certain spatial-ordered pattern with two kinds of spins. 
We defer those ideas to future studies.
 
In summary, we introduced a generic SE scheme applicable to spin-based QBs, 
obtained by inspecting the dynamical equations of motion for spin populations 
[cf. Eq. (\ref{eq:sigz_eom}) for a lattice with nearest-neighbor couplings]. 
Through a modulation of spin energy gap, the engineered QB can harness spin-spin correlations 
to mitigate aging effect in the storage stage, 
thereby achieving an 
advantage in extending the longevity of charged QBs. 
We expect that the resulting scale-invariant protection strategy, 
which is applicable in noisy environments, to play a central role in boosting the performance of QBs.

{\it Acknowledgement.--}
We thank Albert Cabot for insightful discussions and assistance in simulations, and Ilia Khait for critical reading of the manuscript. This work was supported by the Natural Sciences and Engineering Research Council (NSERC) of Canada Discovery Grant and the Canada Research Chairs Program.


%

\clearpage
\renewcommand{\thesection}{\Roman{section}} 
\renewcommand{\thesubsection}{\Alph{subsection}}
\renewcommand{\theequation}{S\arabic{equation}}
\renewcommand{\thefigure}{S\arabic{figure}}
\renewcommand{\thetable}{S\arabic{table}}
\setcounter{equation}{0}  
\setcounter{figure}{0}

\begin{widetext}

{\Large{\bf Supplemental material:} Boosting quantum battery performance by structure engineering}
\\
\\
\\

In this supplemental material, we analyze spin dynamics in the single-excitation sector,
derive coupled equations of motion for spin operators, which govern the dynamics of quantum battery (QB) in the storage phase while taking into account
many-excitation sectors,  and present additional simulation results that complement those included in the main text.

\section{I. Dynamical evolution in the single-excitation sector}

In the main text, we presented simulations for 
$\langle \sigma_j^z(t)\rangle$  and $\langle \sigma_j^+(t)  \sigma_{j'}^-(t)  \rangle$
based on Eqs. (\ref{eq:a7}) and (\ref{eq:a10}) below, respectively. For completeness of our presentation,
we cover here essential results from Ref. \cite{Cabot.19.PRL} 
to explain how these expressions are derived. 

In the single-excitation sector, spin dynamics is fully governed by the master equation 
$\partial \rho_S/\partial t=\mathcal{K}\rho_S=-i(K\rho_S-\rho_SK^{\dagger})$ with $K=H_S-i\sum_j\frac{\gamma_j}{2}\sigma_j^+\sigma_j^-$ a quadratic effective Hamiltonian. 
To solve this master equation, one just needs the eigenspectrum of $K$. 
We place the eigenvalues in the vector $\boldsymbol{\Omega}=(\{\Omega_n\})$ 
and organize the right and left eigenvectors of $K$ in the
matrices $\boldsymbol{M_R}=(\{|K_n\rangle\})$, $\boldsymbol{M_L}=(\{\langle K_n^{\ast}|\})$ respectively,  with $n$ running from 1 to $N$. 
Resorting to the Jordan-Wigner transformation tailored for the single-excitation subspace, one finds the following eigenvalues
\begin{equation}
\boldsymbol{\Omega}~=~(\{\Omega_k^-\},\{\Omega_k^+\})=(\Omega_1,\Omega_2,\Omega_3,\cdots,\Omega_N),
\end{equation}
where $\Omega_k^{\pm}$ is given by \cite{Cabot.19.PRL}
\begin{equation}\label{eq:Ome_pm}
\Omega_k^{\pm}~=~\frac{\Omega_A+\Omega_B}{2}\pm\frac{1}{2}\sqrt{(\Omega_A-\Omega_B)^2+16\lambda^2\cos^2(k/2)}.
\end{equation}
%
\begin{figure}[tbh!]
 \centering
\includegraphics[width=0.5\columnwidth] {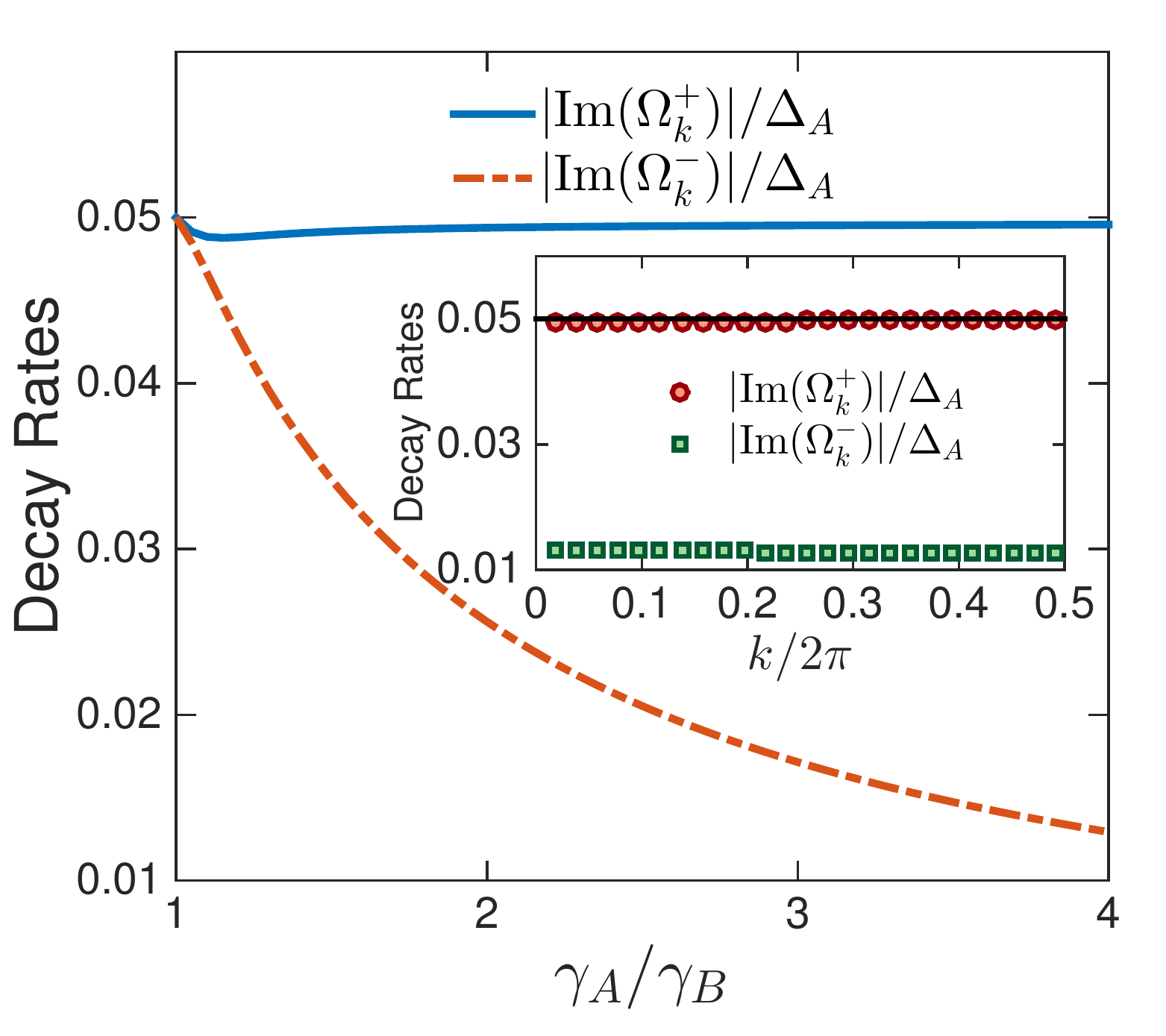}
\caption{Decay rates $|\mathrm{Im}(\Omega_k^{+})|/\Delta_A$ (blue solid line) and $|\mathrm{Im}(\Omega_k^{-})|/\Delta_A$ (red dash-dotted line) of the one-excitation sector with varying ratio $\gamma_A/\gamma_B$ and a fixed $k=20\pi/(N+1)$. 
Inset: Decay rates $|\mathrm{Im}(\Omega_k^{+})|/\Delta_A$ (red circles) and $|\mathrm{Im}(\Omega_k^{-})|/\Delta_A$ (green squares) as a function of index $k=2\pi l/(N+1)$ ($l=1,2,\cdots,N/2$) with $\Delta_B=0.25\Delta_A$. 
The black solid line marks the uniform decay rate $|\mathrm{Im}(\Omega_k^{+})|=|\mathrm{Im}(\Omega_k^{-})|$ for the uniform lattice with $\Delta_B=\Delta_A$. 
Other parameters are $N=50$, $\gamma_{j}/\Delta_{j}=0.05$, $\lambda=0.05\Delta_A$.}
\label{fig:rate}
\end{figure}
%
$\Omega_{A(B)}=\Delta_{A(B)}-i\gamma_{A(B)}$ and $k=2\pi l/(N+1)$ with $l=1,2,\cdots,N/2$. 
The eigenvalues of $K^{\dagger}$ are the complex conjugates $\Omega_k^{\pm\ast}$.
In general, the eigenvalues of the Liouvillian $\mathbb{L}$ can be constructed by using $\Omega_k^{\pm}$ and $\Omega_k^{\pm\ast}$ as detailed in Ref. \cite{Torres.14.PRA}. 
Interestingly, for the single-excitation sector with smallest decay rates, 
the corresponding eigenvalues of the Liouvillian $\mathbb{L}$ are $-i\Omega_k^{\pm}$ and $i\Omega_k^{\pm\ast}$. Hence, the absolute values of the imaginary parts of $\Omega_k^{\pm}$ set the decay rates of the slowest modes. 
As can be seen, for uniform spin lattice with $\gamma_A=\gamma_B\equiv\gamma_0$, we get $\Omega_A=\Omega_B$, 
hence the imaginary parts of $\Omega_k^{\pm}$ coincide, and take the value $\gamma_0$; for uniform QBs, $\gamma_0^{-1}$ 
sets the storage time scale \cite{Pirmoradian.19.PRA}. 
In contrast, for dimeric lattices with $\gamma_A\neq\gamma_B$, the intrinsic decay rates are modified. 
Noting that for vanishing dipole-dipole interactions, $\lambda=0$, we still have two bands provided that $\Omega_A\neq\Omega_B$. 
In Fig. \ref{fig:rate}, we present results for the decay rates $|\mathrm{Im}(\Omega_k^{\pm})|$ (`Im' takes imaginary part hereafter) for the slowest single-excitation sector. From Fig. \ref{fig:rate} and its inset, we clearly observe that the engineered decay rates $|\mathrm{Im}(\Omega_k^{\pm})|$ of two bands in dimeric spin lattices can be made both smaller than that of the uniform limit ($\gamma_A=\gamma_B=0.05\Delta_A$ in present simulations), 
thereby implying a much longer storage time for a dimeric QB, as compared with its non-engineered counterpart.

The corresponding eigenvectors are expressed as \cite{Cabot.19.PRL}
\begin{equation}
\boldsymbol{M_R}~=~(\{v_k^{\prime}|g\rangle\},\{u_k^{\prime}|g\rangle\}),~~\boldsymbol{M_L}~=~(\{\langle g|v_k\},\{\langle g|u_k\}).
\end{equation}
Here, $|g\rangle$ is the global ground state of the spin lattice with $\sigma_j^z|g\rangle=-|g\rangle$. 
The operators $u_k^{\prime},~v_k^{\prime}$ are defined as 
\bea
v_k^{\prime} &=& \sqrt{\frac{4}{N+1}}\sum_{m=1}^{N/2}\Big(\sigma_{2m-1}^+\sin\theta_k\sin[k(m-1/2)]+\sigma_{2m}^+\cos\theta_k\sin(km)\Big),\nonumber\\ 
u_k^{\prime} &=& \sqrt{\frac{4}{N+1}}\sum_{m=1}^{N/2}\Big(\sigma_{2m-1}^+\cos\theta_k\sin[k(m-1/2)]-\sigma_{2m}^+\sin\theta_k\sin(km)\Big).
\eea
Here, $\theta_k$ is determined by $\tan2\theta_k=-4\lambda\cos(k/2)/(\Delta_A-\Delta_B-i(\gamma_A-\gamma_B))$. 
Operators $v_k$ and $u_k$ are obtained from $v_k^{\prime}$ and $u_k^{\prime}$, by replacing spin raising operators with lowering ones, respectively.

In terms of the eigenvectors of the effective Hamiltonian $K$, we introduce the decomposition of the reduced system density matrix
\begin{equation}\label{eq:a4}
\rho_S~=~\left(\sum_{nm}\mathcal{P}_{n,m}+\sum_n\mathcal{P}_{n,0}+\sum_n\mathcal{P}_{0,n}+\mathcal{P}_{0,0}\right)\rho_S.
\end{equation}
Here, $n,~m$ are indices running from 1 to $N$, we denote projectors as 
\bea
\mathcal{P}_{n,m}\rho_S &=& \langle K_n^{\ast}|\rho_S|K_m^{\ast}\rangle|K_n\rangle\langle K_m|,\nonumber\\
\mathcal{P}_{n,0}\rho_S &=& \langle K_n^{\ast}|\rho_S|g\rangle|K_n\rangle\langle g|,\nonumber\\
\mathcal{P}_{0,n}\rho_S &=& \langle g|\rho_S|K_n^{\ast}\rangle|g\rangle\langle K_n|,\nonumber\\
\mathcal{P}_{0,0}\rho_S &=& \langle g|\rho_S|g\rangle|g\rangle\langle g|.
\eea
To illustrate the occurrence of quantum synchronization, we study the ensemble average $\langle \sigma_j^x(t)\rangle$,
\bea
\langle \sigma_j^x(t)\rangle &=& 2\mathrm{Re}\langle \sigma_j^-(t)\rangle~=~2\mathrm{Re}\langle e_j|\rho_S(t)|g\rangle\nonumber\\
&=& 2\mathrm{Re}\langle e_j|\sum_n\mathcal{P}_{n,0}\rho_S(t)|g\rangle. 
\eea
Here, $|e_j\rangle=\sigma_j^+|g\rangle$. It is easy to show that $\mathcal{P}_{n,0}\rho_S(t)=\mathcal{P}_{n,0}\rho_S(0)e^{-i\Omega_nt}$ 
using $\mathcal{K}|K_n\rangle\langle g|=-i\Omega_n|K_n\rangle\langle g|$, hence we have
\begin{equation}\label{eq:a7}
\langle \sigma_j^x(t)\rangle~=~2\mathrm{Re}\left[\sum_{n=1}^NF_{n}(j)e^{-i\Omega_n t}\right],
\end{equation}
with $F_n(j)=\langle K_n^{\ast}|\rho_S(0)|g\rangle\langle e_j|K_n\rangle$ and $\rho_S(0)=|\Phi_0\rangle\langle\Phi_0|$ ($|\Phi_0\rangle$ is the initial state).
For our purpose, we also consider $\langle\sigma_j^z(t)\rangle$ in the single-excitation sector. 
Using Eq. (\ref{eq:a4}) and the relation $\mathcal{P}_{n,m}\rho_S(t)=\mathcal{P}_{n,m}\rho_S(0)e^{-i(\Omega_n-\Omega_m^{\ast})t}$, 
we immediately find that 
\begin{equation}\label{eq:a8}
\langle\sigma_j^z(t)\rangle~=~2\langle e_j|\sum_{n,m}\mathcal{P}_{n,m}\rho_S(t)|e_j\rangle-1~=~2\sum_{n=1,m=1}^NG_{n,m}(j)e^{-i(\Omega_n-\Omega_m^{\ast})t}-1.
\end{equation}
Here, $G_{n,m}(j)=\langle e_j|K_n\rangle\langle K_m|e_j\rangle\langle K_n^{\ast}|\rho_S(0)|K_m^{\ast}\rangle$ with $|K_n^{\ast}\rangle$ and $\langle K_n|$ the right and left eigenvectors of $K^{\dagger}$ with eigenvalues $\Omega_n^{\ast}$. 
Analogously, we obtain the following expression for spin-spin correlations 
$\langle \sigma_j^+(t)\sigma_{j'}^-(t)\rangle$:
\bea
\langle \sigma_j^+(t)\sigma_{j'}^-(t)\rangle &=& \langle e_{j'}|\rho_S(t)|e_j\rangle\nonumber\\
&=& \sum_{n=1,m=1}^NW_{n,m}(j,j')e^{-i(\Omega_n-\Omega_m^{\ast})t}
\label{eq:a10}
\eea
with $W_{n,m}(j,j')=\langle K_n^{\ast}|\rho_S(0)|K_m^{\ast}\rangle\langle K_m|e_j\rangle\langle e_{j'}|K_n\rangle$.

In Fig. \ref{fig:traj_ratio}, we present trajectories $\langle \sigma_j^x(t)\rangle$ using Eq. (\ref{eq:a7}) while varying the frequency ratio $\Delta_B/\Delta_A$. From this comparison, we observe that for relative small ratios $\Delta_B/\Delta_A$ a phase synchronization emerges at the trajectory level as the two trajectories depict almost perfect anti-phase oscillations, while when the ratio $\Delta_B/\Delta_A$ approaches unity, this anti-phase oscillation becomes obscure as can be seen from Fig. \ref{fig:traj_ratio} (d),  consistent with findings in Ref. \cite{Cabot.19.PRL} for short lattices.
%
\begin{figure}[tbh!]
 \centering
\includegraphics[width=0.7\columnwidth] {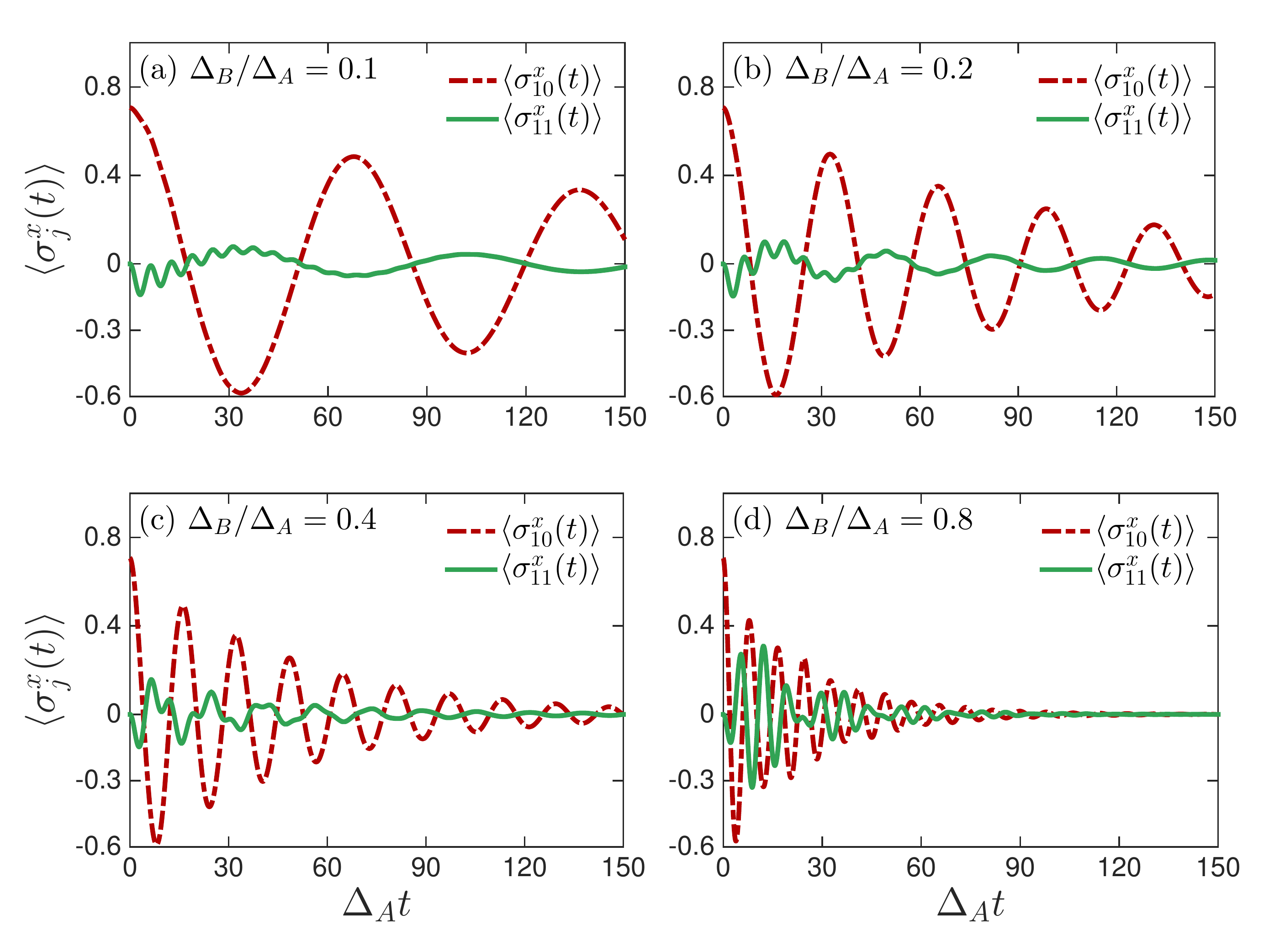}
\caption{Trajectories for $\langle\sigma_{10}^x(t)\rangle$ (red dash-dotted line) and $\langle\sigma_{11}^x(t)\rangle$ (green solid line) 
using Eq. (\ref{eq:a7}) for a dimeric lattice while varying the detuning ratio $\Delta_B/\Delta_A$. 
(a) $\Delta_B/\Delta_A=0.1$, (b) $\Delta_B/\Delta_A=0.2$, (c) $\Delta_B/\Delta_A=0.4$, and (d) $\Delta_B/\Delta_A=0.8$. 
We adopt the initial state $|g\rangle/\sqrt{2}+(|e_{j=10}\rangle+|e_{j=12}\rangle)/2$. 
Other parameters are $N=50$, $\gamma_{j}/\Delta_{j}=0.05$, $\lambda=0.05\Delta_A$.}
\label{fig:traj_ratio}
\end{figure}
\section{II. Many excitations case: Dynamical equations for expectation values of spin operators}

In this section, we first illustrate how to introduce a superradiant decay channel for spins in the discharging phase when the cavity is coupled to the spin chain. 
To this end, we consider the so-called bad cavity limit, $\kappa\gg \eta,~\gamma_j$ \cite{Pirmoradian.19.PRA}. 
In this limit, the cavity degrees of freedom can be adiabatically eliminated using the solution \cite{Holland.14.PRL}
\begin{equation}\label{eq:aa}
a~\simeq~-\frac{2i\eta}{\kappa+i\Delta_A}J_A^{-}-\frac{2i\eta}{\kappa+i\Delta_B}J_B^{-},
\end{equation}
where $J_{A(B)}^-\equiv\sum_{j\in\mathrm{odd}(\mathrm{even})}\sigma_j^-$ are collective lowering spin operators.
Noting that frequency detuning $\Delta_{A,B}=\omega_{A,B}-\omega_c$ can be made small relative to $\kappa$ for bad cavities, 
we  simplify Eq. (\ref{eq:aa}) as $a\simeq-2i\eta J^-/\kappa$ with $J^-=J_A^{-}+J_B^{-}=\sum_j\sigma_j^{-}$, yielding
\bea\label{eq:rho_s}
\frac{\partial}{\partial t}\rho_{S} &=& -i[H_S,\rho_S]+\sum_{j=1}^N\gamma_j\mathcal{L}[\sigma_j^-]\rho_S+\Gamma\mathcal{L}[J^-]\rho_S.
\eea
Here, $H_S=\sum_{j=1}^N\frac{\Delta_j}{2}\sigma_j^z+\sum_{j=1}^{N-1}\lambda(\sigma_j^+\sigma_{j+1}^-+\mathrm{H.c.})$ is the Hamiltonian of dimer spin lattice, $\Gamma=4\eta^2/\kappa$ denotes a Purcell-enhanced emission rate marking a cavity-induced superradiant process, which is designed to be the dominant decay channel for spins in the discharging phase. By setting $\Gamma=0$, we recover the quantum master equation used for analyzing the storage phase in the main text.

To derive dynamical equations from the master equation Eq. (\ref{eq:rho_s}), 
we use the relations $\mathrm{Tr}\{A[H_S,\rho_S]\}=\mathrm{Tr}\{[A,H_S]\rho_S\}$ and
\begin{equation}
\mathrm{Tr}\{A\mathcal{L}[O]\rho_S\}~=~\frac{1}{2}\mathrm{Tr}\{[O^{\dagger},A]O\rho_S\}+\frac{1}{2}\mathrm{Tr}\{\rho_SO^{\dagger}[A,O]\}
\end{equation}
for an arbitrary spin operator $A$ and a Lindblad superoperator 
$\mathcal{L}[O]\rho_{S}=(2O\rho_{S}O^{\dagger}-O^{\dagger}O\rho_{S}-\rho_{S}O^{\dagger}O)/2$.
The dynamical equation for $\langle \sigma_j^z\rangle$ takes the form (explicit time dependence is suppressed hereafter):
\bea\label{eq:b2}
\frac{d}{dt}\langle \sigma_j^z\rangle &=& -2i\lambda\left[\langle \sigma_j^+\sigma_{j+1}^-\rangle+\langle\sigma_j^+\sigma_{j-1}^-\rangle-\langle\sigma_{j+1}^+\sigma_j^-\rangle-\langle\sigma_{j-1}^+\sigma_j^-\rangle\right]-\gamma_j\langle 1+\sigma_j^z\rangle\nonumber\\
&&-\Gamma\langle 1+\sigma_j^z\rangle-\Gamma\sum_{m\neq j}\Big[\langle \sigma_j^+\sigma_m^-\rangle+\mathrm{c.c.}\Big],
\eea
where `c.c' denotes complex conjugate. 
From the above equation of motion, it is evident that the discharging phase always benefits from a superradiant decay channel with a Purcell-enhanced emission rate, $\Gamma\gg\gamma_{A,B}$, thereby achieving a fast discharging process \cite{Pirmoradian.19.PRA}. We highlight that our setup lacks a permutation symmetry, that is, spin correlations depend on the indices due to the presence of a nearest-neighbor dipole-dipole coupling, in contrast to the scenario studied in Ref. \cite{Pirmoradian.19.PRA}. 

The above dynamical equation should be solved subject to those for $\langle\sigma_n^+\sigma_m^-\rangle$. Below we limit our attention to the storage phase when $\Gamma$ is tuned to zero.
 For $\langle\sigma_n^+\sigma_m^-\rangle$, we find it convenient to separately treat two scenarios, due to the presence of nearest-neighbor dipole-dipole coupling: (i) $m=n\pm1$ and (ii) $m\neq n, n\pm1$,
\begin{itemize}
\item Case (i):
\bea\label{eq:b3}
\frac{d}{dt}\langle\sigma_n^+\sigma_{n\pm1}^-\rangle &=& \left[i(\Delta_n-\Delta_{n\pm1})-\frac{\gamma_n+\gamma_{n\pm1}}{2}\right]\langle\sigma_n^+\sigma_{n\pm1}^-\rangle\nonumber\\
&&-i\frac{\lambda}{2}\Big[\langle\sigma_n^z\rangle-\langle\sigma_{n\pm1}^z\rangle\Big]-i\lambda\Big[\langle\sigma_n^z\sigma_{n\mp1}^+\sigma_{n\pm1}^-\rangle-\langle\sigma_{n\pm1}^z\sigma_n^+\sigma_{n\pm2}^-\rangle\Big].
\eea
\item Case (ii):
\bea\label{eq:b4}
\frac{d}{dt}\langle\sigma_n^+\sigma_m^-\rangle &=& 
\Big[i(\Delta_n-\Delta_m)-\frac{\gamma_n+\gamma_m}{2}\Big]\langle\sigma_n^+\sigma_m^-\rangle\nonumber\\
&&-i\lambda\sum_{h=n\pm1}\langle\sigma_n^z\sigma_{m}^-\sigma_{h}^+\rangle+i\lambda\sum_{h=m\pm1}\langle\sigma_m^z\sigma_{n}^+\sigma_{h}^-\rangle.
\eea
\end{itemize}
To form a closed set of coupled dynamical equations, we adopt a semiclassical cumulant approximation that is applicable to large spin numbers: correlations are expanded to second order. 
Particularly, here we approximate $\langle\sigma_n^z\sigma_m^+\sigma_h^-\rangle\approx\langle\sigma_n^z\rangle\langle\sigma_m^+\sigma_h^-\rangle$ \cite{Meiser.10.PRAa}. 
By doing so, we neglect correlations of the type 
$\langle\sigma_n^z\sigma_m^+\rangle$ ($n\neq m$) \cite{Holland.14.PRL}. 
Accordingly, we  approximate Eqs. (\ref{eq:b3}) and (\ref{eq:b4}) as
\begin{itemize}
\item Case (i):
\bea\label{eq:b5}
\frac{d}{dt}\langle\sigma_n^+\sigma_{n\pm1}^-\rangle &\approx& \left[i(\Delta_n-\Delta_{n\pm1})-\frac{\gamma_n+\gamma_{n\pm1}}{2}\right]\langle\sigma_n^+\sigma_{n\pm1}^-\rangle-i\frac{\lambda}{2}\Big[\langle\sigma_n^z\rangle-\langle\sigma_{n\pm1}^z\rangle\Big]\nonumber\\
&&-i\lambda\Big[\langle\sigma_n^z\rangle\langle\sigma_{n\mp1}^+\sigma_{n\pm1}^-\rangle-\langle\sigma_{n\pm1}^z\rangle\langle\sigma_n^+\sigma_{n\pm2}^-\rangle\Big].
\eea
\item Case (ii):
\bea\label{eq:b6}
\frac{d}{dt}\langle\sigma_n^+\sigma_m^-\rangle &\approx& \left[i(\Delta_n-\Delta_m)-\frac{\gamma_n+\gamma_m}{2}\right]\langle\sigma_n^+\sigma_m^-\rangle\nonumber\\
&&-i\lambda\sum_{h=n\pm1}\langle\sigma_n^z\rangle\langle\sigma_h^+\sigma_m^-\rangle+i\lambda\sum_{h=m\pm1}\langle\sigma_m^z\rangle\langle\sigma_n^+\sigma_h^-\rangle.
\eea
\end{itemize}
Eqs. (\ref{eq:b2}), (\ref{eq:b5}) and (\ref{eq:b6}) form a closed set, which can be numerically propagated by means of, for instance, Runge-Kutta algorithm subject to the open boundary condition for the spin lattice.

\section{III. Additional simulation results}
In this section, we include additional simulations that complement those shown in the main text. 
For characterizing the performance of QBs in the storage phase, one can also look at the normalized population defined as
\begin{equation}
\mathcal{P}(t)~=~\frac{\sum_{j=1}^N\frac{\Delta_j}{2}\langle \sigma_j^z(t)\rangle}{\sum_{j=1}^N\Delta_j/2}.
\end{equation}
Similarly to the main text, we denote by $\mathcal{P}_{d(u)}(t)$ the population of the dimeric (uniform) spin lattice. 
A typical set of results is presented in Fig. \ref{fig:pop}. 
%
\begin{figure}[tbh!]
 \centering
\includegraphics[width=0.47\columnwidth] {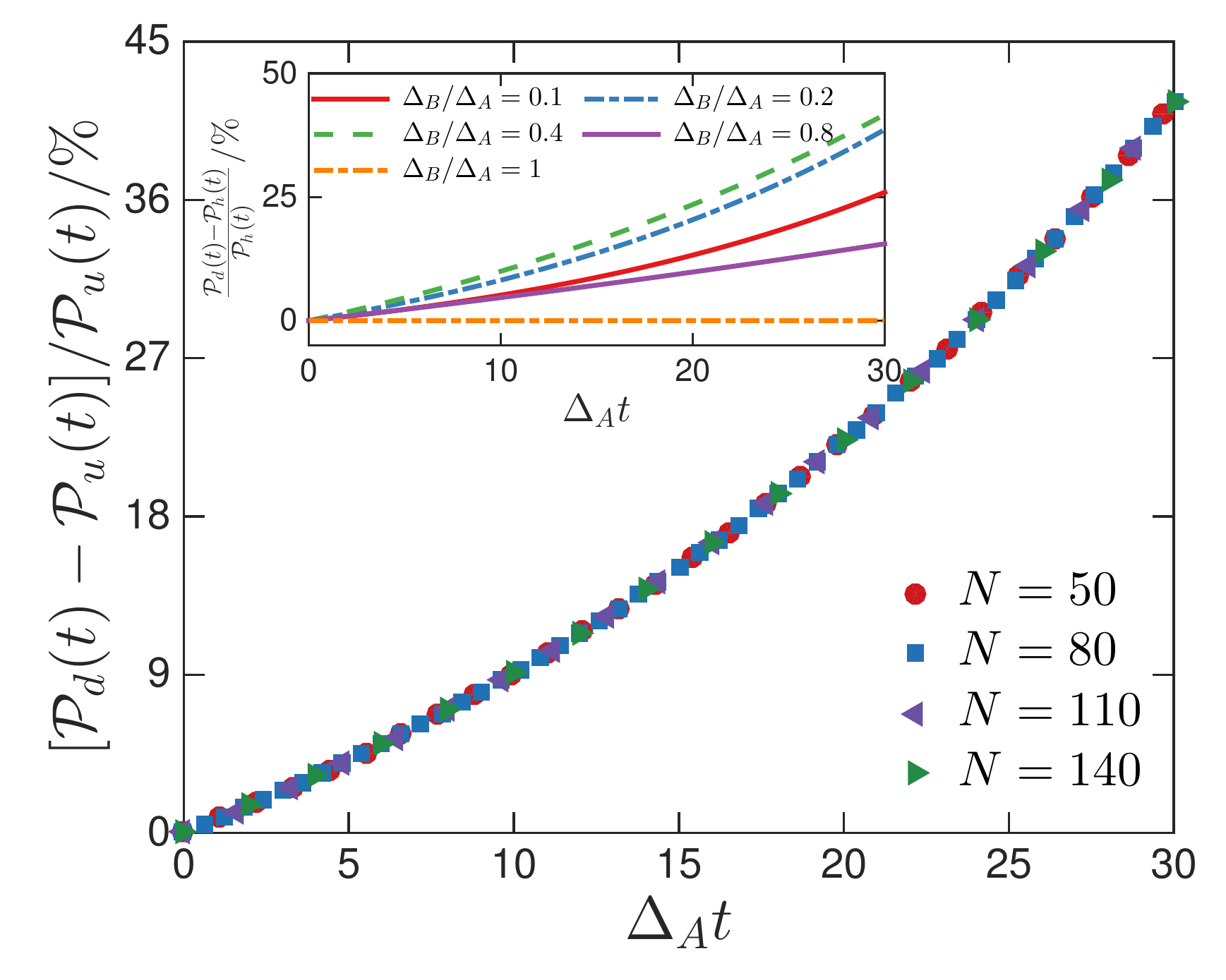}
\caption{Dynamics of relative excess population $[\mathcal{P}_d(t)-\mathcal{P}_u(t)]/\mathcal{P}_u(t)$ in the storage phase as a function of spin number $N$ with a fixed detuning ratio $\Delta_B/\Delta_A=0.25$. The inset shows the relative excess energy when varying the detuning ratio $\Delta_B/\Delta_A$ and a fixed number of spins $N=80$.
 Other parameters are $\gamma_{A(B)}/\Delta_{A(B)}=0.05$ and $\lambda=0.05\Delta_A$.}
\label{fig:pop}
\end{figure}
As can be seen, the normalized population depicts almost the same behavior as the normalized energy shown in the main text.
Therefore, the normalized population can also serve as a figure of merit for characterizing QBs in the storage phase.

In Fig. \ref{fig:long_time}, we check the long time behavior of the absolute excess energy $\mathcal{E}_d(t)-\mathcal{E}_h(t)$.
%
\begin{figure}[tbh!]
 \centering
\includegraphics[width=0.5\columnwidth] {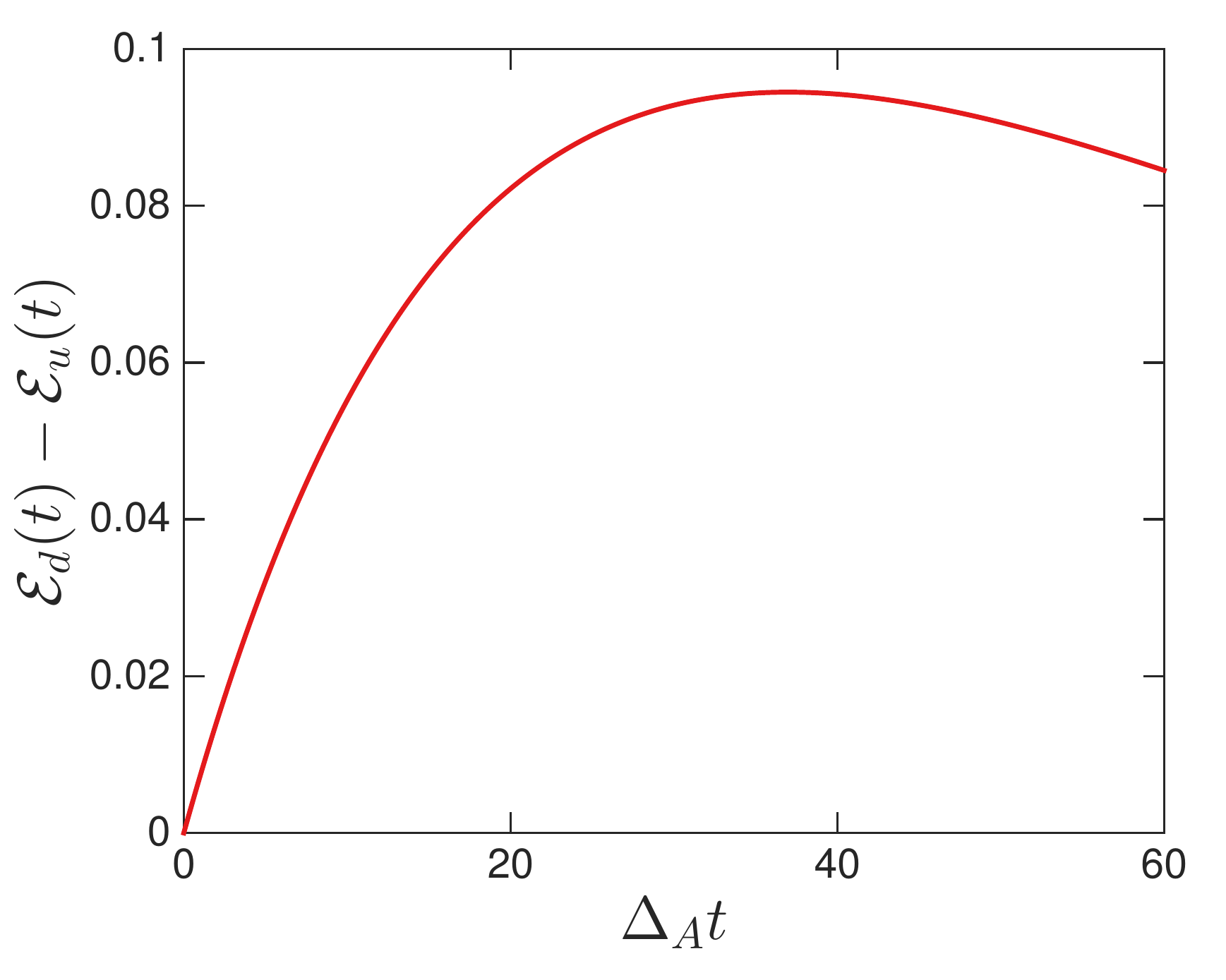}
\caption{Dynamics of excess energy density $\mathcal{E}_d(t)-\mathcal{E}_h(t)$ in the storage phase for a dimeric battery with $\Delta_B/\Delta_A=0.25$. Other parameters are $N=80$, $\gamma_{j}/\Delta_{j}=0.05$, $\lambda=0.05\Delta_A$.}
\label{fig:long_time}
\end{figure}
We find that $\mathcal{E}_d(t)-\mathcal{E}_h(t)$ depicts a turnover behavior when increasing the storage time, indicating that there is an optimal protection time for engineered QBs.

In Fig. \ref{fig:loglog}, we analyze the dependence of $[\mathcal{E}_d(t)-\mathcal{E}_u(t)]/\mathcal{E}_u(t)$ on time.
We observe a power-law behavior, $\mathcal{E}_d(t)/\mathcal{E}_u(t)\propto t^{\alpha}$,
with $\alpha=1.5$ at longer times.
%
\begin{figure}[tbh!]
 \centering
\includegraphics[width=0.5\columnwidth] {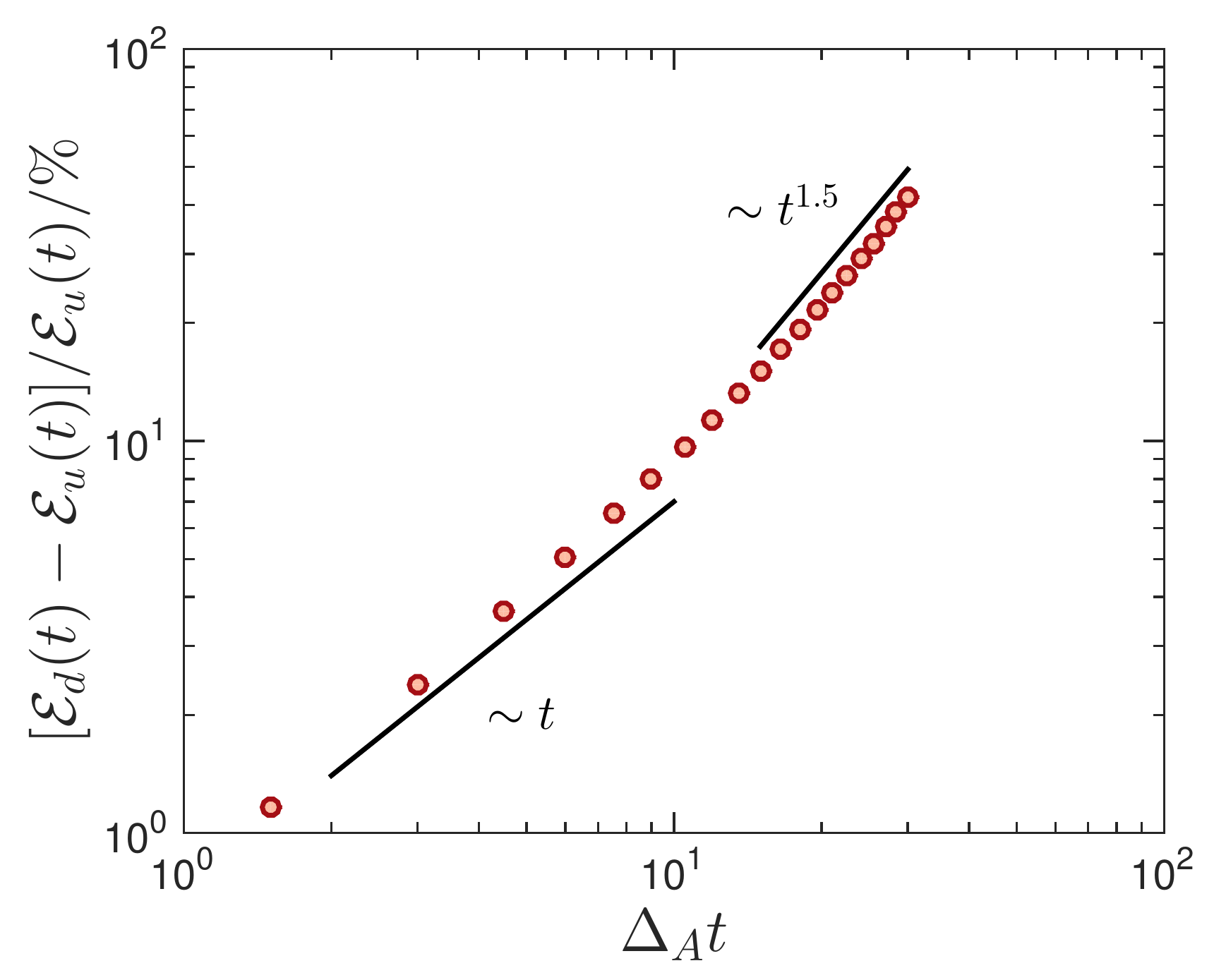}
\caption{Dynamics of the relative stored energy excess $[\mathcal{E}_d(t)-\mathcal{E}_u(t)]/\mathcal{E}_u(t)$ as a function of time with a fixed detuning ratio $\Delta_B/\Delta_A=0.25$. 
Other parameters are $N=80$, $\gamma_{A(B)}/\Delta_{A(B)}=0.05$ and $\lambda=0.05\Delta_A$.}
\label{fig:loglog}
\end{figure}

\end{widetext}

\end{document}